\documentclass[a4paper,11pt]{article}

\usepackage[utf8]{inputenc}
\usepackage[T1]{fontenc}
\usepackage{times}
\usepackage{fullpage}
\usepackage{graphicx}
\usepackage{float}
\usepackage{authblk}

\usepackage{hyperref}
\hypersetup{
    colorlinks=true,
    linkcolor=black,
    citecolor=black,
    filecolor=blue,
    urlcolor=blue,
    hyperfootnotes=false
}

\usepackage{natbib}
\bibpunct{(}{)}{;}{a}{,}{,}
\newcommand*{\doi}[1]{\href{http://dx.doi.org/\detokenize{#1}}{\detokenize{#1}}}

\title{\textbf{Realistic simulations reveal extensive sample-specificity of {RNA}-seq biases}}
\author{Botond Sipos}
\author{Greg Slodkowicz}
\author{Tim Massingham\footnote{Current address: Oxford Nanopore Technologies, Edmund Cartwright House, 4 Robert Robinson Avenue, Oxford Science Park, Oxford, OX4 4GA, United Kingdom}}
\author{Nick Goldman}
\affil{\footnotesize \textit{European Molecular Biology Laboratory, European Bioinformatics Institute (EMBL-EBI), Wellcome Trust Genome Campus, Hinxton, Cambridge CB10 1SD, United Kingdom}}
\date{}

\begin{document}
\maketitle
\vspace{-1.5cm}
\begin{center}
\href{http://creativecommons.org/licenses/by-nc-sa/3.0/}{\includegraphics[scale=0.45]{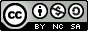}}\footnote{This work is licensed under a \href{http://creativecommons.org/licenses/by-nc-sa/3.0/}{Attribution-NonCommercial-ShareAlike 3.0 Unported License}.}
\end{center}

\textbf{
In line with the importance of RNA-seq, the bioinformatics community has produced numerous data analysis tools incorporating methods to correct sample-specific biases. However, few advanced simulation tools exist to enable benchmarking of competing correction methods. We introduce the first framework to reproduce the properties of individual RNA-seq runs and, by applying it on several datasets, we demonstrate the importance of accounting for sample-specificity in realistic simulations.
}

Studying the repertoire of RNA molecules in the cell (transcriptomics) is one of the most important fields of modern biology, underlying basic research into biological processes and countless clinical and practical applications. High-throughput sequencing of RNA (RNA-seq) is a relatively new addition to the transcriptomics toolbox \citep{wang09, marguerat10} with many   implementations for different sequencing platforms. Likely due to the wide availability of the platform, protocols based on Illumina sequencing dominate the field and considering the relatively cheap coverage offered --- a feature that is key when measuring the magnitude of transcript expression --- will surely remain fundamental assays in transcriptomics.

The advantages of RNA-seq over previous methods are manifold and include wider dynamic range, better reproducibility and single base resolution \citep{wang09,marguerat10}. However, RNA-seq library preparation and sequencing necessarily introduce biases \citep{sendler11} with the potential to distort the biological conclusions. Evidence for the impact of biases in standard Illumina RNA-seq is overwhelming. Two of them are documented to affect fragment coverage over the transcripts: random hexamer priming leads to dependence on the sequence context around fragment start/end \citep{hansen10} while PCR biases introduce a dependence on fragment GC content \citep{speed12}. 

Priming biases have been claimed to be reproducible \citep{hansen10}; however, the evidence is compelling that GC-dependent biases are sample-specific \citep{risso11,speed12}. Sample-specific biases can have a striking effect on coverage trends, making them more dependent on the laboratory of source than the tissue of origin as uncovered by a large scale exploratory study \citep{gelfand12}. Consequently, it is not surprising that recent studies find a strong effect of GC-content \citep{hansen12,risso11} when testing for differential expression, a key study design for uncovering new biology through transcriptomics. 

The bioinformatics and biostatistics communities have produced numerous tools and methods for RNA-seq analysis \citep{pachter11, alamancos13}, from simple counting of coverage of annotated features \citep{htseq} to advanced tools aiming to jointly reconstruct the transcriptome and quantify expression \citep{guttman10,trapnell10}, or to estimate isoform expression levels in presence of multi-mapping reads \citep{turro11}. Many quantification \citep{roberts11, glaus12} and normalization \citep{risso11,hansen12} tools incorporate approaches to learn the biases and correct for them. 

Nevertheless, these approaches are not necessarily suitable for deepening our understanding of the data due to the gap between their relative simplicity and the complex nature of biological biases. Simulation is traditionally an important strategy for modelling data, testing hypotheses and evaluating inference methods, widely exploited in all fields of computational biology. The evaluation of RNA-seq bias correction methods, however, has been based mostly on a limited number of ``gold standard’’ quantitative PCR replicates or cross-platform comparisons. The only other option has been Flux Simulator \citep{griebel12}, which has pioneered the realistic simulation of RNA-seq experiments.

We present the \textit{rlsim} package (\href{https://github.com/sbotond/rlsim}{https://github.com/sbotond/rlsim}), the first advanced simulation framework to reproduce the properties of specific Illumina RNA-seq datasets. \textit{rlsim} simulates key steps of RNA-seq library construction protocols (e.g.~fragmentation, priming, PCR amplification, size selection) with particular focus on the latter steps (PCR, size selection) that can be informed by the analysis of specific datasets. To this end, the package provides tools for estimating insert size distribution, corrected and uncorrected relative expression levels and GC-dependent amplification efficiencies using an approach that can be thought of as the extension of the method of \citet{speed12} to RNA-seq data and is similar to bias correction approaches such as that implemented in \textit{BitSeq} \citep{glaus12}. These parameters, estimated from real datasets, are combined with a ``priming affinity’’ model inspired by a thermodynamical model of oligonucleotide hybridization. The details of the framework are described in the package documentation (\href{http://bit.ly/rlsim-doc}{http://bit.ly/rlsim-doc}).

In order to to survey the contribution simulation can make to understanding priming and PCR biases, we have analyzed 13 paired-end RNA-seq datasets produced by the standard random priming Illumina protocol and four datasets produced by the leading dUTP-based \citep{levin10} strand-specific protocol. The datasets were produced by 9 different laboratories, and all were derived from different biological samples.

To quantify the fit (``realism’’) of the simulation model and thus the contribution of the \textit{rlsim}’s advanced modeling of sample-specific GC-dependent efficiencies and priming affinity,
we use a strategy akin to model selection by Approximate Bayesian Computation \citep{sunnaker13}. 
Briefly, we estimate parameters from the reads mapped to the respective Ensembl canonical transcriptomes and simulate datasets under two settings: one the ``full’’ model, making use of the maximum amount of knowledge available about the dataset, and a second, ``flat’’ one using uncorrected expression levels, fixed GC-independent amplification efficiencies, uniform priming affinities and no simulated polyadenylation. To evaluate model fit, we calculate the Spearman rank correlation of appropriate summary statistics in the simulated and real datasets. We also establish the expected upper bound of correlations by analysing corresponding features in single-ended library construction technical replicates \citep{bullard10}. The full analysis pipeline with results is available at GitHub (\href{https://github.com/sbotond/paper-rlsim}{https://github.com/sbotond/paper-rlsim}).

The first finding of our study is that the GC-dependent bias profiles do not necessarily have a symmetric, unimodal shape centered around a GC-content of 50\% (see \textit{effest\_report.pdf} listed under the \textit{Bias/} directory [\href{http://bit.ly/rlr-bias}{http://bit.ly/rlr-bias}] in the GitHub repository, along with other reports mentioned), as has previously been assumed based on analysis of DNA sequencing data \cite{speed12}. Some strand-specific datasets (e.g.~SRR549333) seem particularly prone to strong and complex biases, likely due to the additional steps in the stranded protocol.

The correlations of various statistics summarizing the properties of datasets indicate that the advanced \textit{rlsim} features indeed help in reproducing real sample biases biases (\textbf{Fig.~1}).  The magnitude of the improvement is, however, highly variable across different datasets, suggestive of sample-specific effects.

\begin{figure}[H]                                                                                                      
\centering                                                                                                             
\includegraphics[scale=0.75]{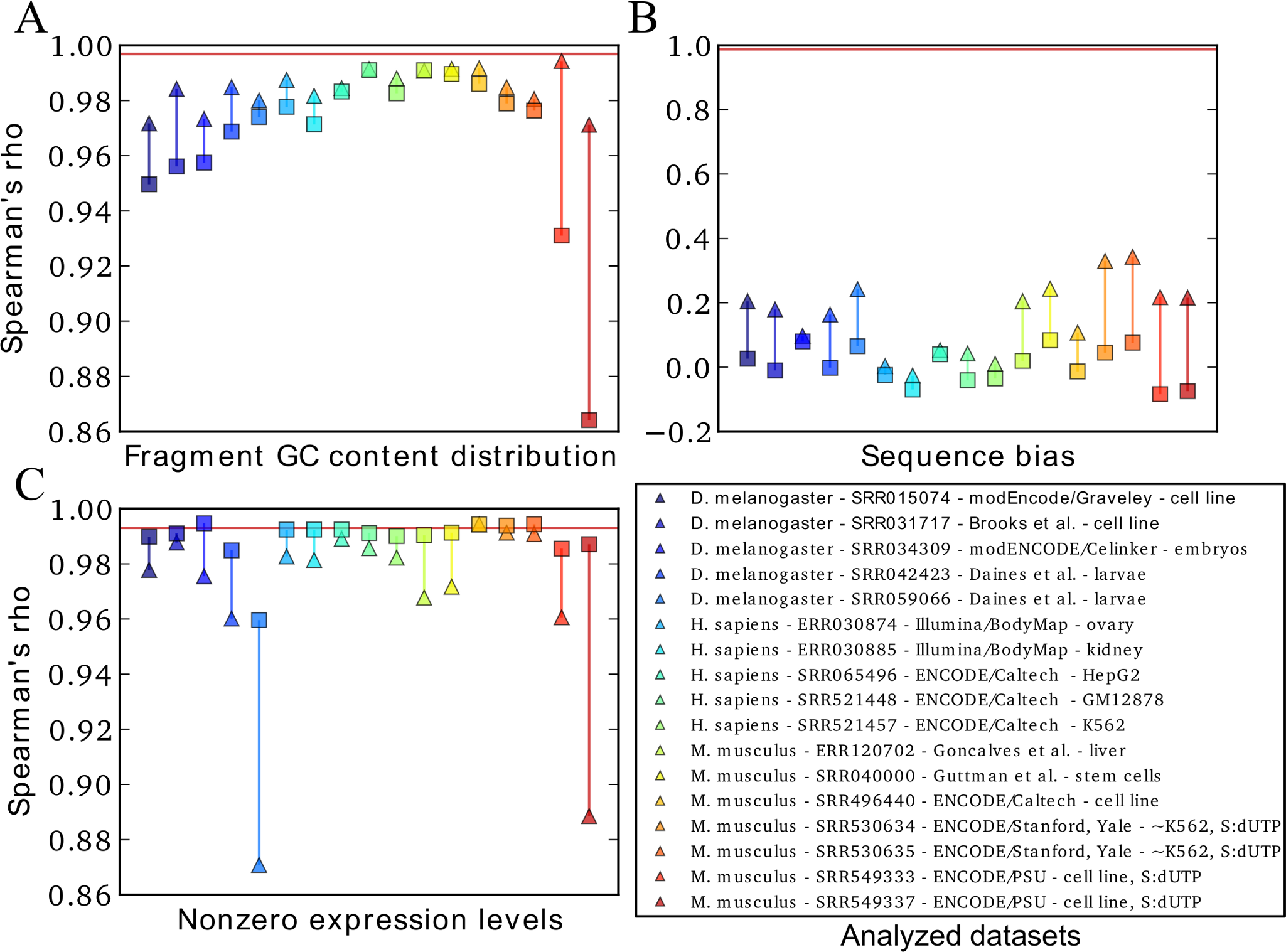}
\caption{
\textbf{\footnotesize Shift in Spearman rank correlation of different summary statistics between real data and data simulated under the ``flat’’ (squares) and ``full’’ (triangles) simulation settings. Horizontal red lines mark the mean rank correlation of corresponding statistics calculated from pairs of single-ended technical replicates by \citet{bullard10}.} \textbf{A.}~Fragment GC content distribution. \textbf{B.}~Mean correlation of base counts near start/end of fragments with a window size of 10. \textbf{C.}~Relative expression levels.
}
\end{figure}

The correlation of the fragment GC content distributions is higher (or essentially the same) in full simulations (\textbf{Fig.~1A}), however it is already very high even in the simpler ``flat’’ simulations, indicating that this feature is mainly determined by the relative expression levels in most datasets.

\citet{hansen10} concluded that despite being caused by the use of random hexamers, priming biases are not explained well by binding energies. The mean correlation of base counts around fragment start and end (\textbf{Fig.~1B}), a feature summarizing sequence-specific biases, is low in the flat simulations and increased in all full simulations, indicating that \textit{rlsim}’s advanced features help reproducing sequence biases. However, the fact that the correlation remains modest ($\sim$0.2) indicates that there are major unaccounted factors shaping this bias and reinforces the conclusions of \citet{hansen10}.

For some transcripts the correlation between coverage and GC content is apparent by visual examination (\textbf{Fig.~2A}; also the \textit{cov\_cmp.pdf} and \textit{cov\_cmp\_flat.pdf}  files linked under \textit{Bias/}). In general, the distribution of transcript fragment coverage correlations showed a mode near zero in all flat simulations, which is not surprising considering the highly stochastic nature of this feature. In the full simulations they are shifted towards one (see \textit{meta\_cov\_cmp.pdf}), slightly in some datasets (ERR030885) and markedly in others (SRR549333).

\begin{figure}[H]                                                                                                      
\centering                                                                                                             
\includegraphics[scale=0.75]{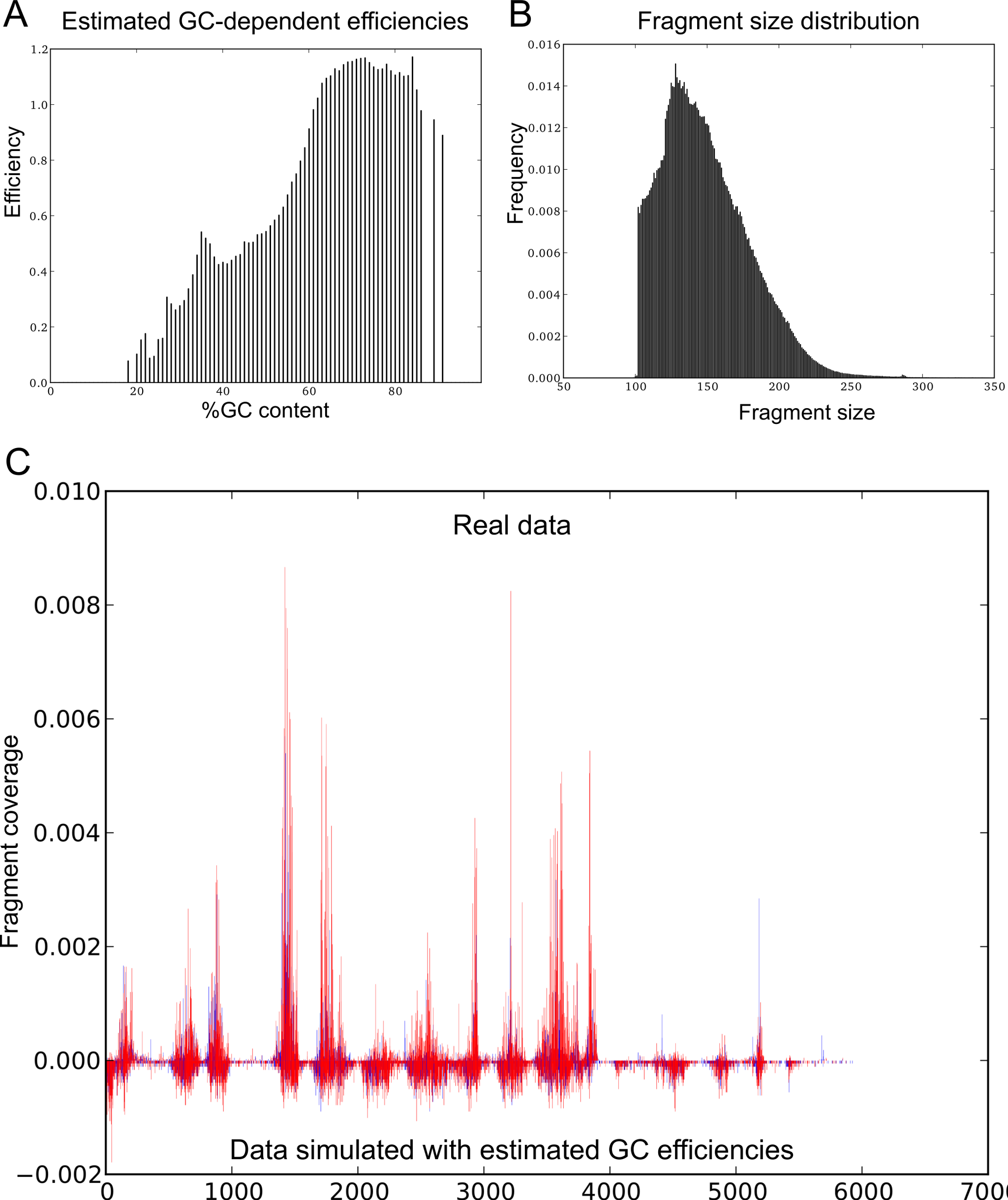}
\caption{
\textbf{\footnotesize
Correlation of local GC content and coverage in real and simulated data in the case of the single spliceform (ENSMUST00000098789) transcribed from the gene Zc3h4 in the mouse RNA-seq dataset SRR549337.} \textbf{A.}~Estimated GC-dependent efficiencies (inferred efficiencies > 1 are rounded to 1 for simulations). \textbf{B.}~Estimated fragment size distribution. \textbf{C.}~The real (positive $y$-domain) and simulated (negative $y$-domain) relative coverage of 5$^\prime$ fragment ends across the transcript (Spearman rank correlation: 0.52), colored by the identity of the respective base (AT: blue, GC: red).
}
\end{figure}

The improvement provided by sample-specific estimates is reinforced by the positive correlation between the magnitude of increase in all statistics and the strength of estimated GC-biases. Also, with the exception of sequence bias, correlations between simulated and real data are almost as high as correlations between library construction-level technical replicates (horizontal lines on \textbf{Fig.~1}). One possible explanation of the slightly poorer correlations is that our simulations ignore splicing events by simulating only the canonical transcriptome.

In summary, our results demonstrate that the additional factors simulated in \textit{rlsim}’s full model help in reproducing the biases in the analyzed data sets. However, the correlation of relative expression levels decreased in the full simulations (\textbf{Fig.~1C}), likely due to the \textit{ad hoc} GC-content correction and the increased complexity of the model. It is, however, important to point out that this does not invalidate the advantages of \textit{rlsim} in benchmarking settings, as the true expression levels are known and the introduced biases are tunable by the user. 

This study can be regarded as the first quantification and implementation of the sum our knowledge about the Illumina RNA-seq data generation process, as summarized by our ability to reproduce relevant features of the data. In line with estimation-based studies \citep{risso11,roberts11,speed12,gelfand12}, our results confirm that RNA-seq biases are highly sample-specific, underlining the importance of considering this in estimation and simulation settings alike. To facilitate benchmarking based on existing data, we have made corrected expression levels and parameters estimated from the analysed datasets available in a separate repository (\href{https://github.com/sbotond/rlsim-params}{https://github.com/sbotond/rlsim-params}). We believe the development and assessment of bias correction methods will remain important in the future and we hope that benchmarking them through fair and controllable simulations will become an important part of the best practices.

\bibliography{paper-rlsim}{}
\end{document}